\documentclass[runningheads]{llncs}
\usepackage[T1]{fontenc}
\usepackage{cite}
\usepackage{graphicx}
\usepackage{subcaption}
\usepackage{adjustbox}
\usepackage{booktabs}
\usepackage{algorithm}
\usepackage{algorithmic}
\usepackage{listings}

\begin{document}
%

\title{GeoNimbus: A serverless framework to build earth observation and environmental services}
\titlerunning{GeoNimbus: A serverless framework to build EOS}
%
\author{Dante D. Sánchez-Gallegos\inst{1}\orcidID{0000-0003-0944-9341} \and Diana Carrizales-Espinoza\inst{2}\orcidID{0000-0002-3925-031X} \and Alejandro Zequeira\inst{2} \and Catherine Torres-Charles\inst{1} \and
J. L. Gonzalez-Compean\inst{2}\orcidID{0000-0002-2160-4407} \and Jesus Carretero\inst{1}\orcidID{0000-0002-1413-4793}}
\authorrunning{Sánchez-Gallegos et al.}
%
\institute{Universidad Carlos III de Madrid, Leganés, Madrid, Spain \\ \email{dantsanc@pa.uc3m.es} \and
Cinvestav Tamaulipas, Cd. Victoria, Tamaulipas, México \\
\email{joseluis.gonzales@cinvestav.mx}}
\maketitle              
\begin{abstract}
Cloud computing has become a popular solution for organizations implementing Earth Observation Systems (EOS). However, this produces a dependency on provider resources. Moreover, managing and executing tasks and data in these environments are challenges that commonly arise when building an EOS. This paper presents GeoNimbus, a serverless framework for composing and deploying spatio-temporal EOS on multiple infrastructures, e.g., on-premise resources and public or private clouds. This framework organizes EOS tasks as functions and automatically manages their deployment, invocation, scalability, and monitoring in the cloud. GeoNimbus framework enables organizations to reuse and share available functions to compose multiple EOS. We use this framework to implement EOS as a service for conducting a case study focused on measuring water resource changes in a lake in the south of Mexico. The experimental evaluation revealed the feasibility and efficiency of using GeoNimbus to build different earth observation studies.

\keywords{earth observation services \and serverless computing \and e-science \and big data \and environmental indexes}
\end{abstract}

\section{Introduction}

Serverless computing has become an alternative for organizations to deploy their systems \cite{ueckermann2020podpac}. Instead of users managing infrastructure details, in this paradigm, the users only submit their functions to a cloud platform \cite{jonas2019cloud}. The serverless platform automatically and transparently manages the execution, replication, scale, and invocation of functions sent by users. This paradigm is suitable for researchers and organizations who want to build Earth Observation Systems (EOS) with multiple interconnected applications. For instance, an EOS based on imagery processing typically could consider three primary modules: extracting, preparing, and processing satellite images. Following this example, organizations could define multiple functions to extract imagery and process them using different preprocessing functions (e.g., radiometric, atmospheric, or cloud detection) and analyze them using indexes for multiple observation subjects such as vegetation, water, fires, and urban expansion, to name a few \cite{ueckermann2020podpac}. 

In these platforms, organizations delegate control over the application deployment and data management to a cloud provider \cite{kurniawan2019introduction}, which could establish a dependency on the infrastructure provider (vendor lock-in) \cite{opara2014critical}. Recently, solutions like Globus Compute \cite{chard2020funcx} have made different efforts to make serverless computing available in different infrastructures instead of relying on a single cloud provider. However, there are still two complex aspects that organizations must consider when adopting in-house serverless. The first one is data movement from user devices to the endpoints. The second one is managing the scale and performance of the functions included in EOS and coordinating dataflows through different infrastructures. These are critical aspects as the analysis processing in earth observation systems essentially is a big data problem where EOS processes large amounts of data \cite{barron2024gis}.

In this paper, we present GeoNimbus\footnote{We choose this name because Geo comes from geospatial, and Nimbus is a type of precipitation cloud.} a serverless framework for creating and deploying spatio-temporal earth observation services (EOS) on multiple infrastructures (e.g., any of on-premise resources, public, or private clouds). We design GeoNimbus to accomplish two goals: the first is to provide organizations and developers with a design-driven framework for creating EOS deployed through multiple infrastructures. The second one is to reduce the complexity of the EOS design so that researchers can conduct environmental studies with a framework that automatically manages the deployment and execution of EOS.

The contributions of GeoNimbus are:
\begin{itemize}
    \item \textit{An in-house function-as-a-service (iH-FaaS) and container-as-a-service (CaaS) approach for EOS} to  automatically handle the deployment and coupling of  applications. 
     \item \textit{A wide-area storage system to handle data exchange through the distributed functions}. This system stores raw data and results of an EOS to make them available to end-users (researchers). 
     \item \textit{Implicit parallelism structures to improve the performance of EOS.}  These structures enhance the efficiency of applications of an EOS by using an implicit parallelism model. 
\end{itemize}

We evaluate GeoNimbus in a case study focused on creating an EOS to perform environmental studies of changes in water resources in Cuitzeo Lake, located in the Mexican state of  Michoacán. We conducted this spatio-temporal study by processing \textit{LandSat8} images from 2013 to 2022. The performance evaluation revealed the usability of our framework for deploying EOS in serverless environments. 

The rest of the paper is as follows: Section \ref{sec:relatedwork} describes the related work; Section \ref{sec:desing} presents the design principles of GeoNimbus; Section \ref{sec:arqnProt} presents the GeoNimbus architecture; Section \ref{sec:casestudy} presents the experimental evaluation; and, Section \ref{sec:conclusions} presents the main conclusion of this work and future work.

\section{Related work} \label{sec:relatedwork}
This section presents the main work on serverless computing and earth observation systems.

Serverless computing provides an easy-to-use framework for deploying and managing services on remote infrastructure. Some examples of serverless platforms are AWS Lambda \cite{lambda},  Google Cloud Functions \cite{cloudfunctions}, and Azure Functions \cite{kurniawan2019introduction}.

Alternative solutions exist in the literature to extend serverless computing to on-premise resources like personal computers and HPC clusters. Globus Compute \cite{chard2020funcx} is a tool that enables developers to convert computers into serverless endpoints to deploy functions-as-a-service (FaaS), allowing them to take advantage of the available resources. For example, developers can deploy functions and services near the data sources to reduce the latency between data and processing resources \cite{ullah2023orchestration}. Nevertheless, Globus Compute requires that users manage the movement of input data to the endpoint where they deployed functions, which is complex when they must process large volumes of data. Users can deploy their functions on the same data storage infrastructure to alleviate this issue.

The framework proposed in this paper is similar to Globus Compute in terms of enabling the deployment of serverless services through multiple infrastructures. However, it differs because GeoNimbus composes various functions into a single system. This allows users and organizations to create automatic dataflow through different functions and creates different design patterns like pipelines and workflows.


Earth observation is a research area where it requires the processing of large volumes of data to perform spatio-temporal studios \cite{nativi2015big}. 
Different authors have proposed approaches for implementing earth observation services using serverless platforms in this context. For example, Kaiser et al. proposed a framework to transform traditional earth observation applications into systems capable of being deployed on serverless platforms \cite{kaiser2021towards}. Ueckermann et al. presented a framework to design earth observation services using serverless resources \cite{ueckermann2020podpac}. Nevertheless, the authors intended these solutions to work in a single cloud provider, which again could produce a vendor lock-in dependency. Furthermore, the cloud's centralization of data and computation could produce efficiency and latency issues. Thus, we propose deploying these serverless solutions through multiple infrastructures to avoid dependencies with public cloud providers and mitigate efficiency and confidentiality issues that could emerge when working in public cloud environments.

\section{GeoNimbus: Design principles} \label{sec:desing}

In this section, we describe the design principles of GeoNimbus, a serverless framework for building and deploying spatio-temporal Earth Observation Systems (EOS). 

\subsection{Automatic composition of EOS dataflows deployed on multiple infrastructures}

\begin{figure}[t]
    \centering
    \includegraphics[width=\linewidth]{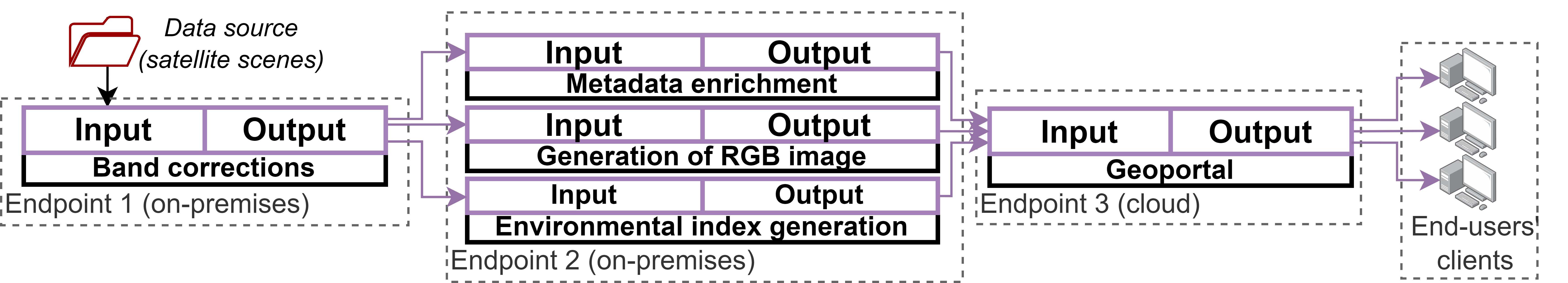}
    \caption{example of an earth observation system (EOS) built with GeoNimbus.}
    \label{fig:exampleflow}
\end{figure}

GeoNimbus enables organizations to create EOS by coupling multiple stages (functions and applications) and deploying them through multiple infrastructures. Figure \ref{fig:exampleflow} shows an EOS built using the framework GeoNimbus. As can be observed, in GeoNimbus, a stage contains I/O interfaces and the code of the function/application. An EOS like the depicted in Figure \ref{fig:exampleflow} is built in three main phases: \textit{i)} design, \textit{ii)} deployment, and \textit{iii)} execution.

First, at design time, a developer creates a configuration file by declaring the stages of a system. These stages contain a function or an application encapsulated into a virtual container. Functions are more suitable for two scenarios: applications that only run for a short period or those deployed on devices with low computation characteristics (e.g., edge devices or personal computers). Organizations could use virtual containers for applications that process large amounts of data through high-performance computers. Furthermore, this configuration file contains the endpoints to deploy the stages and the interconnections between the stages. These interconnections define the execution order of the stages in a system and the data dependencies between these stages.

Next, at deployment time, the GeoNimbus framework considers a controller that deploys the stages over the infrastructure specified by designers. A daemon previously installed on that infrastructure deployment performs this deployment process. This daemon establishes a connection with the controller to couple the stages by following the interconnections defined at design time. These interconnections create data connectors that use the file system, memory, or network resources. File system and memory channels are more suitable for moving data between stages at the same endpoint, whereas the network channels are suitable for moving data through different endpoints. GeoNimbus provides these data connectors, and developers are responsible for choosing the appropriate data connector to fulfil their requirements. In the next section, we describe these data management mechanisms in detail.
At this point, developers can build an EOS of different stages interconnected through multiple infrastructures.

Finally, at execution time, the stages start processing data arriving at their input interfaces. Then, the stages send their results to the next stage through their output stage. 

Figure \ref{fig:exampleflow} shows an example of  EOS built by using GeoNimbus, where the \textit{band corrections} stage processes data from a data source and delivers its outputs to the \textit{Metadata enrichment}, \textit{Generation of RGB image}, and \textit{Environment index generation} stages. Subsequently, these stages deliver their outputs to a \textit{Geoportal} deployed on a cloud instance.

\subsection{Movement of data through multiple infrastructures} \label{sec:storage}

\begin{figure}[t]
    \centering
    \includegraphics[width=\linewidth]{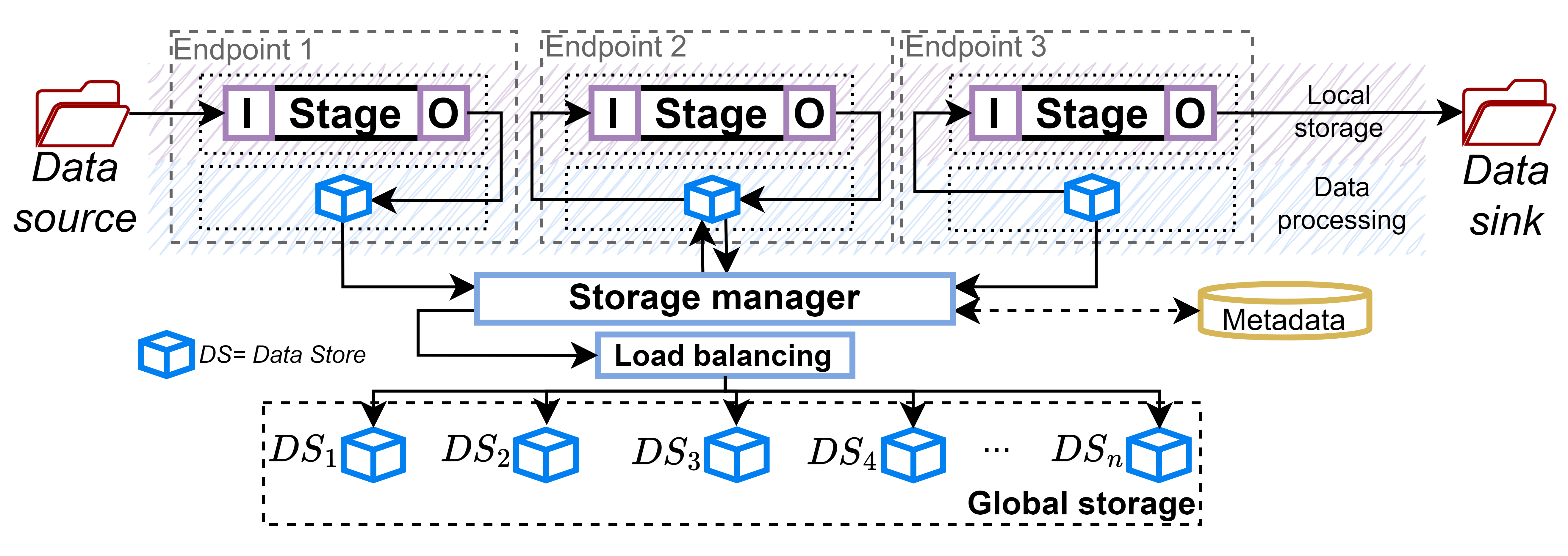}
    \caption{Wide area storage system to support the exchange of data through applications in an EOS.}
    \label{fig:datamovement}
\end{figure}

GeoNimbus performs the management of data through stages deployed on different endpoints using a wide-area storage system \cite{CARRIZALESESPINOZA2024}. Figure \ref{fig:datamovement} shows the conceptual representation of this storage system, which includes artifacts called \textit{data stores} (DS). Moreover, the storage system implements a storage manager, which implements a load balancing service based on a utilization factor metric \cite{morales2018data} and a metadata service to record the location of data stored in the DS. A DS is a virtual storage unit that stores the data required and produced by stages (their applications) in a system. In GeoNimbus, we classify DS into two types: local and global.

Organizations deploy a local DS on the same endpoints where the data is processed. Thus, it implements interfaces to allow stages to push and pull data. For example, a stage performs a pull operation to retrieve its input data, whereas a push operation loads its results into a data store. When a stage performs a pull operation, the DS also registers the data on the storage manager's metadata service. Furthermore, this process also creates a catalog that other stages can subscribe to access this data. This enables the transference of data between two stages of a system by following the following flow:
\begin{enumerate}
    \item A stage $A$, deployed on an endpoint A, pushes its results to a local DS, also deployed on the same endpoint.
    \item The DS $A$ publishes a catalog in the metadata service of the storage manager, and the DS $B$ subscribes to this catalog.
    \item The data is moved from DS $A$  to the DS $B$. The DSs use the storage manager as an intermediary to perform this transference. 
    \item The DS $B$ writes the received data in the input directory (I) of the consumer stage, and the daemon on the endpoint invokes the execution of the application in the stage for data processing.
\end{enumerate}

This process is repeated through each interconnection of stages in the system until the last stage delivers the results to a data sink. This automatically creates a continuous data flow through the stages in the system, even if an organization deploys them through multiple infrastructures (endpoints).

Global data stores preserve data for long periods and make them accessible to other users. Thus, global data stores create a global storage service, which could be composed of virtual storage instances on the cloud or in large storage clusters. The distribution of data through the available global DS is performed using an algorithm based on the utilization factor of the storage instances, as described by Morales et al. \cite{morales2018data}. Note that the system administrator decides when to publish data into the global storage service, keeping control of the data held by the system owners. Furthermore, this global storage allows organizations to connect multiple EOS to share data and create more complex systems. 

\subsection{Implicit parallelism using parallel patterns} \label{sec:autoscaling}

In GeoNimbus, designers can create parallel patterns to improve the performance of their solutions. These patterns are based on the manager/worker pattern \cite{sanchez2022puzzlemesh}, where a manager distributes a set of contents through replicas of the same stage (workers). The workers process the data and deliver their outputs to the next stage or a storage location like a DS. Designers can configure the number of workers in a parallel pattern during design time. Furthermore, GeoNimbus implements an autoscaling scheme to avoid bottlenecks in an EOS due to input workload changes or data delivery delays.

\begin{figure}[t]
    \centering
    \includegraphics[width=.9\linewidth]{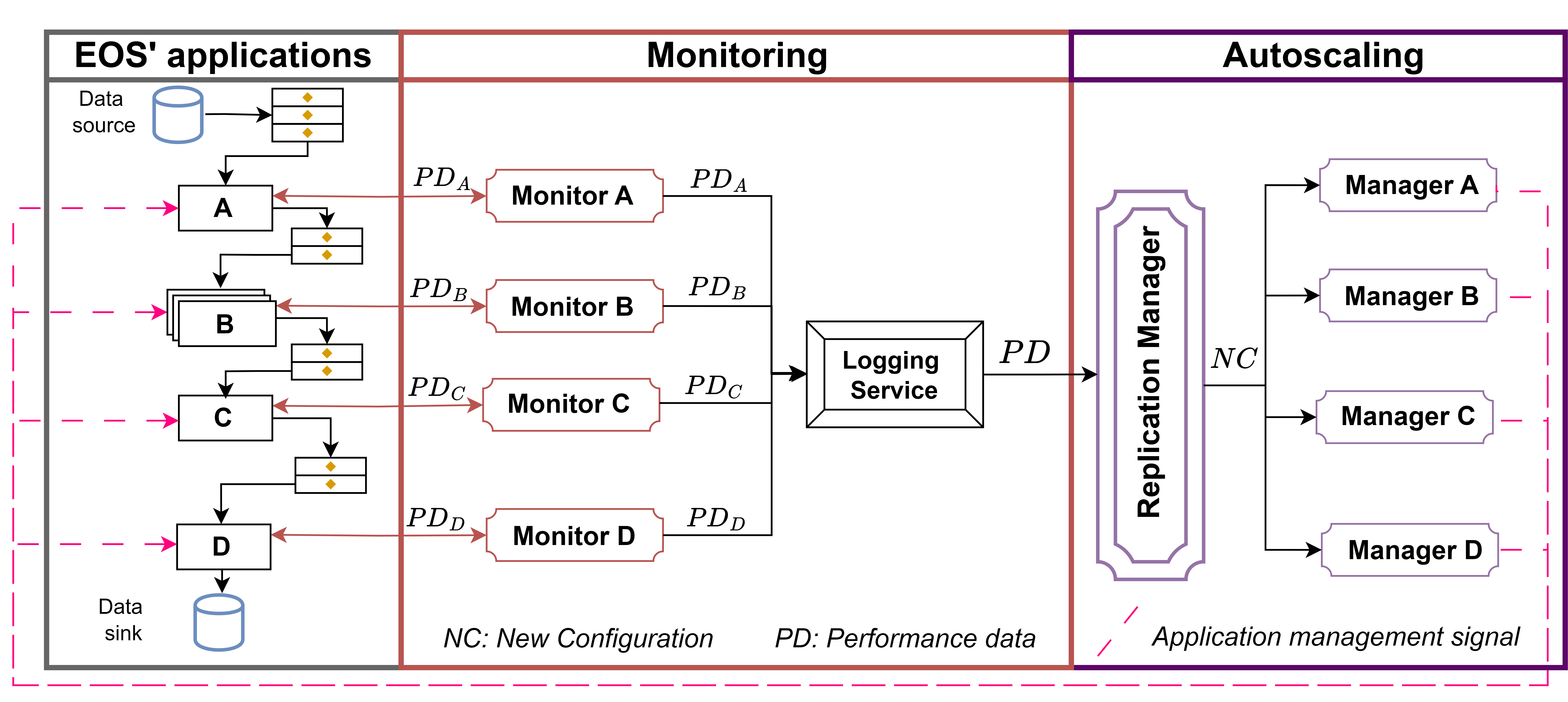}
    \caption{Autoscaling scheme designed to manage the number of workers in a pattern.}
    \label{fig:autoscaling}
\end{figure}

Figure \ref{fig:autoscaling} shows the conceptual representation of the autoscaling scheme. As can be seen, it includes a set of monitors that are deployed on the endpoints where the stages (A, B, C, and D) are running. For each stage, monitors collect their service time, the amount of data processed, and the average waiting time of each task (input content) to be processed. These measurements ($PD_x$) are sent to a logging service, which a replication manager consumes. 

The replication manager obtains the throughput of each application and identifies that \textit{stages} with the lowest throughput that can potentially produce bottlenecks, as follows:  $Btl = \mathrm{MINTHPOS}(thpApps)$, where $thpApps$ is the throughput of each stage.

Once a bottleneck ($Btl$) is identified, the next step is adding a new worker to this stage to increase its performance. Adding new workers does not always ensure the improvement of the performance of an application; it depends on the available physical resources (number of cores), the management of these resources by the application, and the resource consumption of other applications. Thus, we first limit the maximum number of workers to the number of physical cores at the endpoint. In a second instance, we continuously monitor the performance of stages and determine if their performance is degraded when new workers are added; thus, these are removed from the pattern. As can be seen in Figure \ref{fig:autoscaling}, the replication manager gives a new configuration ($NC$) to the indicated stage manager that was identified with a bottleneck ($Btl$). In this example, stage B needs workers; the state manager sends a signal to its corresponding state to create the indicated workers for its application.

\section{Architecture and prototype implementation} \label{sec:arqnProt}
\begin{figure}[t]
    \centering
    \includegraphics[width=.9\linewidth]{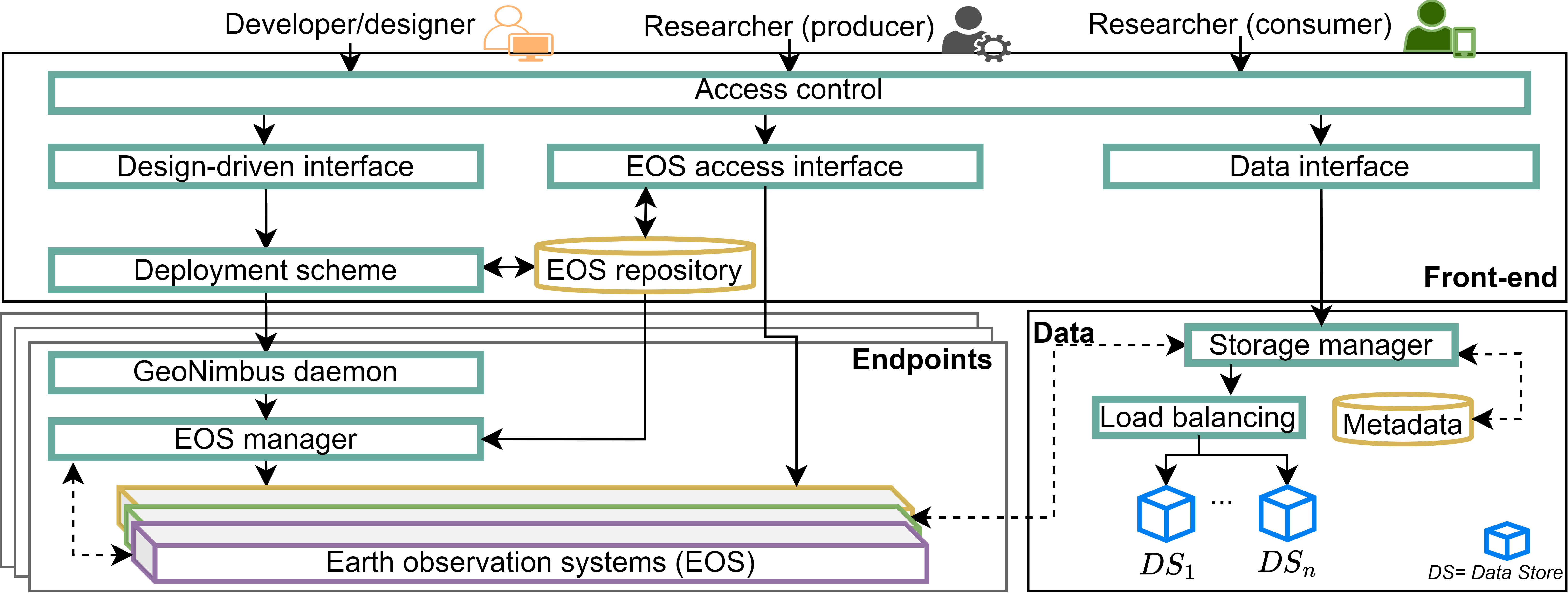}
    \caption{The GeoNimbus architecture.}
    \label{fig:archi}
\end{figure}

Figure \ref{fig:archi} depicts GeoNimbus' architecture, composed of three main layers: control front-end, endpoints, and data. The \textit{front-end} layer includes access control services to manage users' access to the rest of the services. Moreover, three interfaces to \textit{i)} design new EOS and deploy them on the endpoints, \textit{ii)} access existing EOS, and \textit{iii)} access the products EOS produces, or raw data loaded into data stores ($DS_n$).

The second layer is the \textit{endpoints} executing EOS. Thus, this layer includes the GeoNimbus daemon to manage the deployment of EOS by creating the stages (containers and functions) required in each system. Moreover, it consists of an EOS manager implementing the parallelism scheme described in Section \ref{sec:autoscaling} to perform the monitoring of the applications in the system and mitigation of bottlenecks. Finally, the \textit{data} layer implements the wide-area storage system described in Section \ref{sec:storage}. Thus, this layer implements the storage manager and global data stores to orchestrate data movement through the different applications in an EOS. 

We implemented a prototype based on GeoNimbus to perform an evaluation based on a case study to analyze changes in vegetation and water resources in Mexico. The autoscaling and applications orchestrator are mainly implemented in C++ language. The data storage system based on data stores is implemented in Python 3.10. The configuration file used to create a system is based on a construction model called PuzzleMesh \cite{sanchez2022puzzlemesh}.

\section{Case study: services to analyze changes in water resources in Mexico} \label{sec:casestudy}

\begin{figure}[t]
    \centering
    \includegraphics[width=\linewidth]{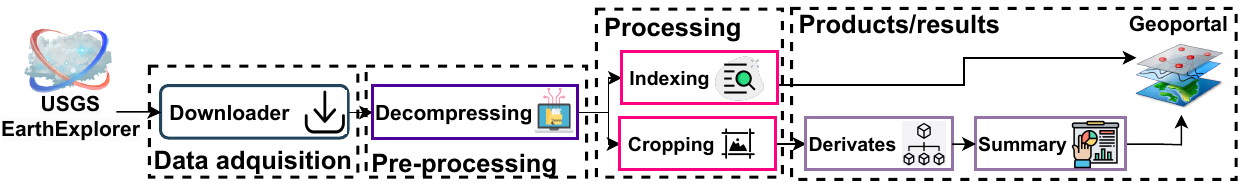}
    \caption{Design of the system used in this case study.}
    \label{fig:casestudy}
\end{figure}

We conduct an evaluation based on a case study of GeoNimbus. This case study is based on a system for processing LandSat8 images and generating normalized difference water indexes (NDWI) \cite{yan2017analysis} to analyze changes in water resources in Lake Cuitzeo. This lake is located in the Mexican state of Michoacán, and in 2005, it used to be the second-largest freshwater lake in Mexico. 

\begin{figure}[t]
    \centering
    \begin{subfigure}[b]{0.32\textwidth}
         \centering
         \includegraphics[width=\textwidth]{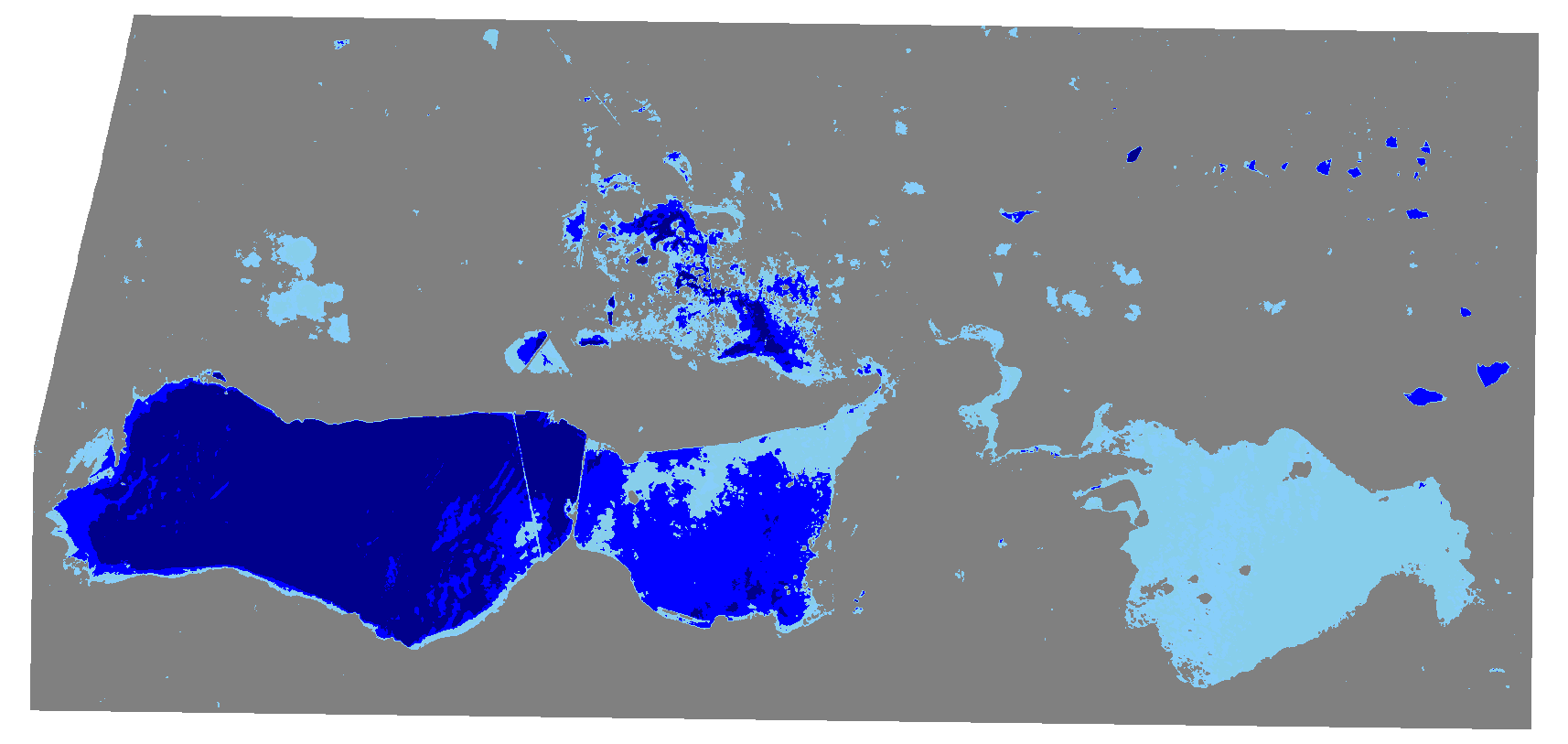}
         \caption{2013.}
         \label{fig:img1}
     \end{subfigure}
     \hfill
     \begin{subfigure}[b]{0.32\textwidth}
         \centering
         \includegraphics[width=\textwidth]{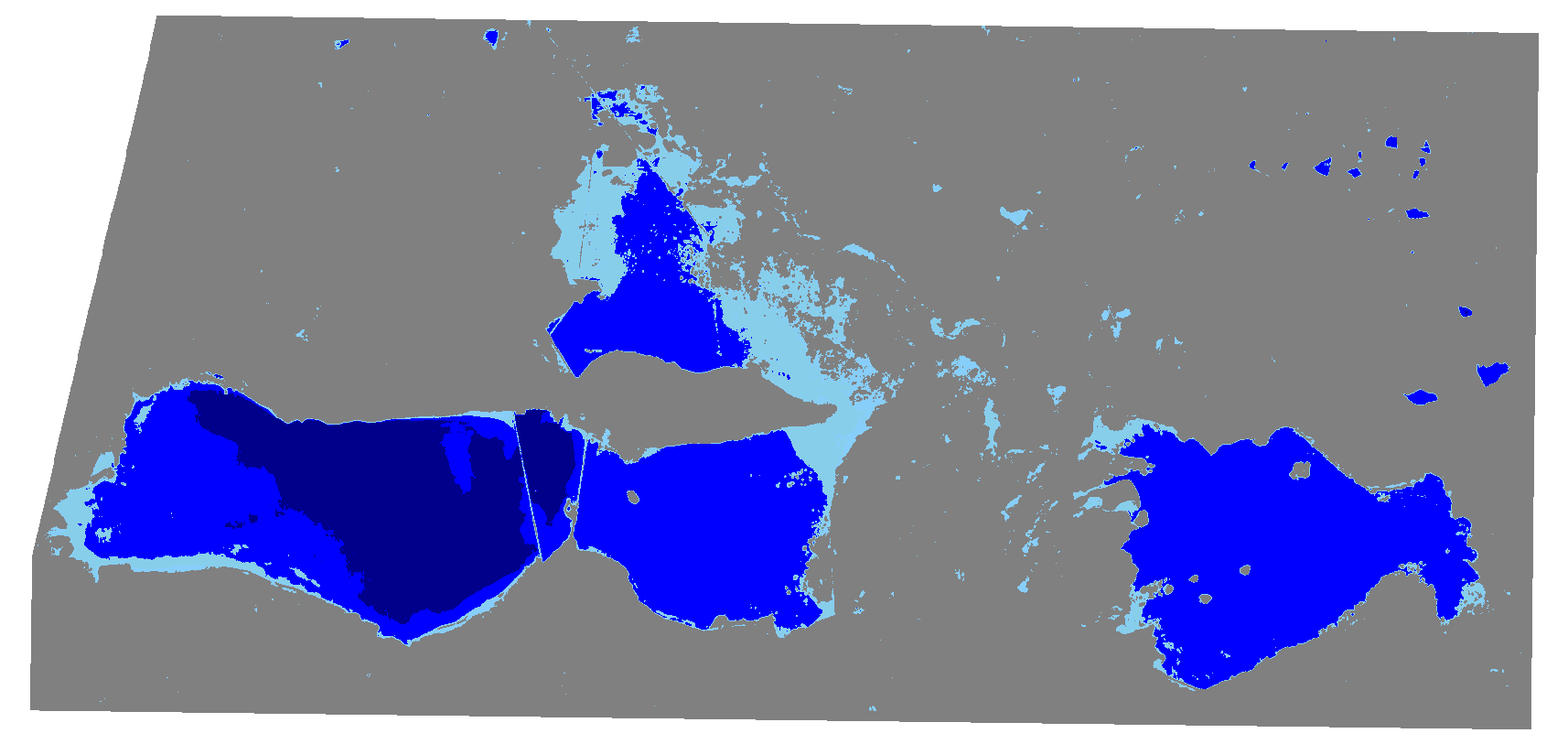}
         \caption{2015.}
         \label{fig:img2}
     \end{subfigure}
     \begin{subfigure}[b]{0.32\textwidth}
         \centering
         \includegraphics[width=\textwidth]{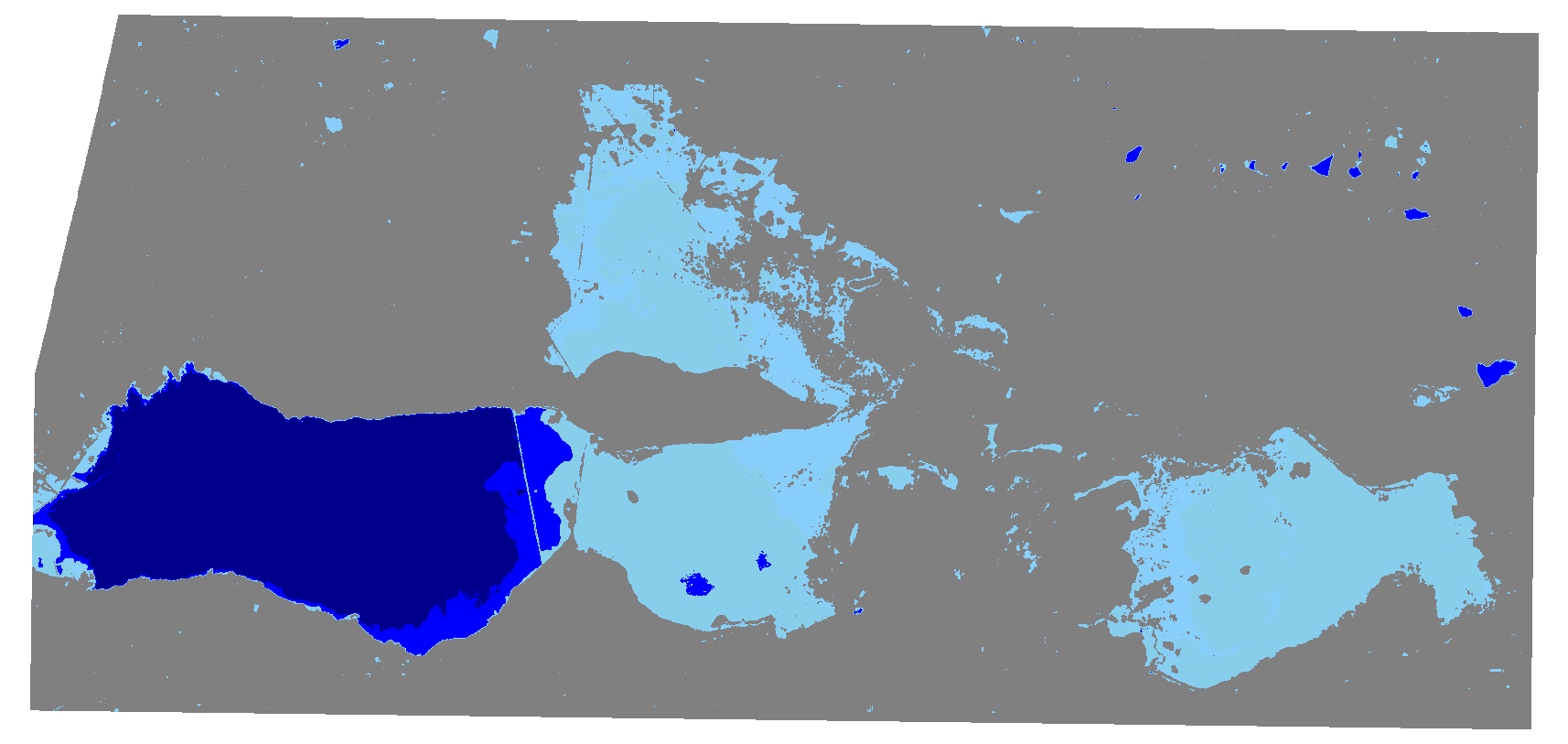}
         \caption{2018.}
         \label{fig:img3}
     \end{subfigure}
     \begin{subfigure}[b]{0.32\textwidth}
         \centering
         \includegraphics[width=\textwidth]{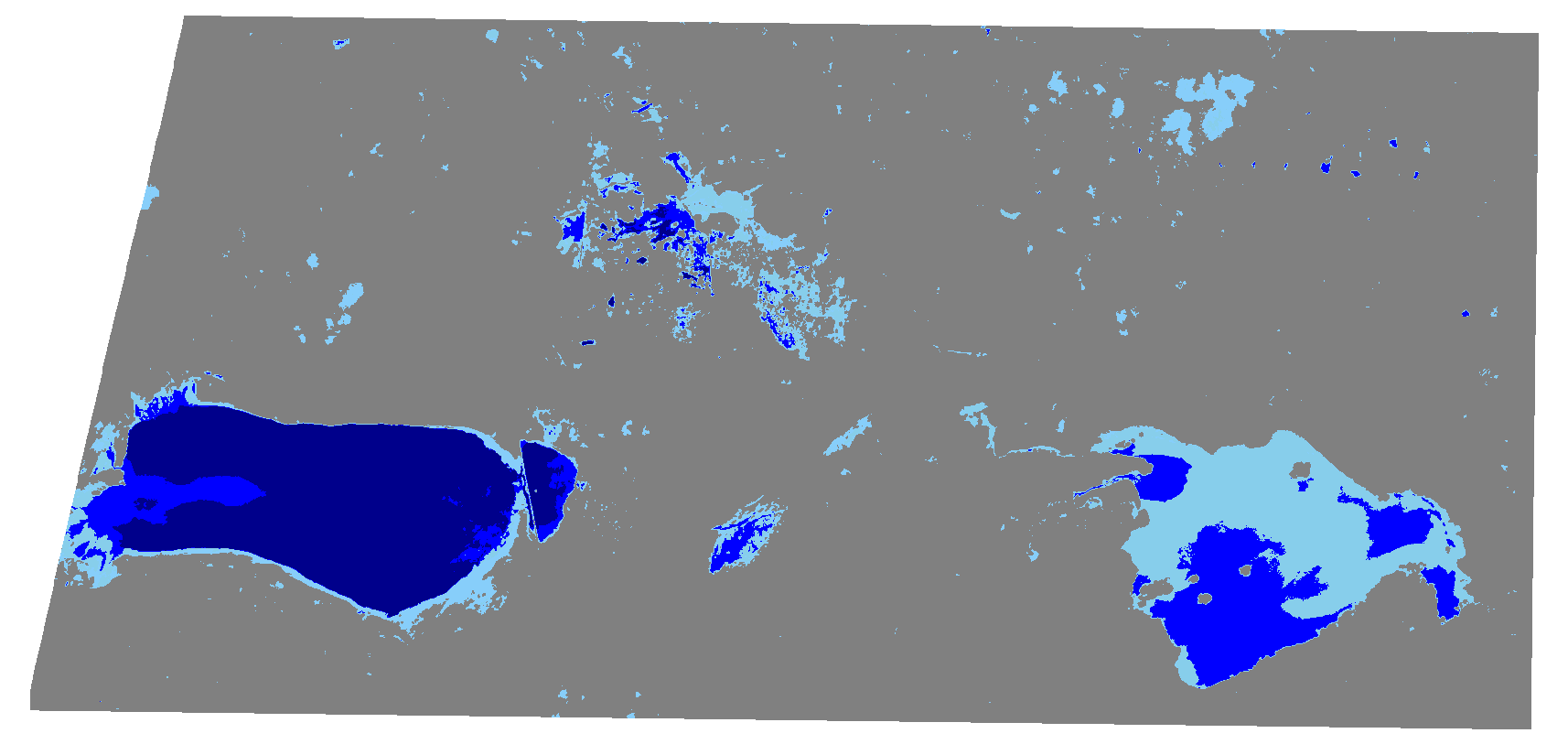}
         \caption{2021.}
         \label{fig:img4}
     \end{subfigure}
     \begin{subfigure}[b]{0.32\textwidth}
         \centering
         \includegraphics[width=\textwidth]{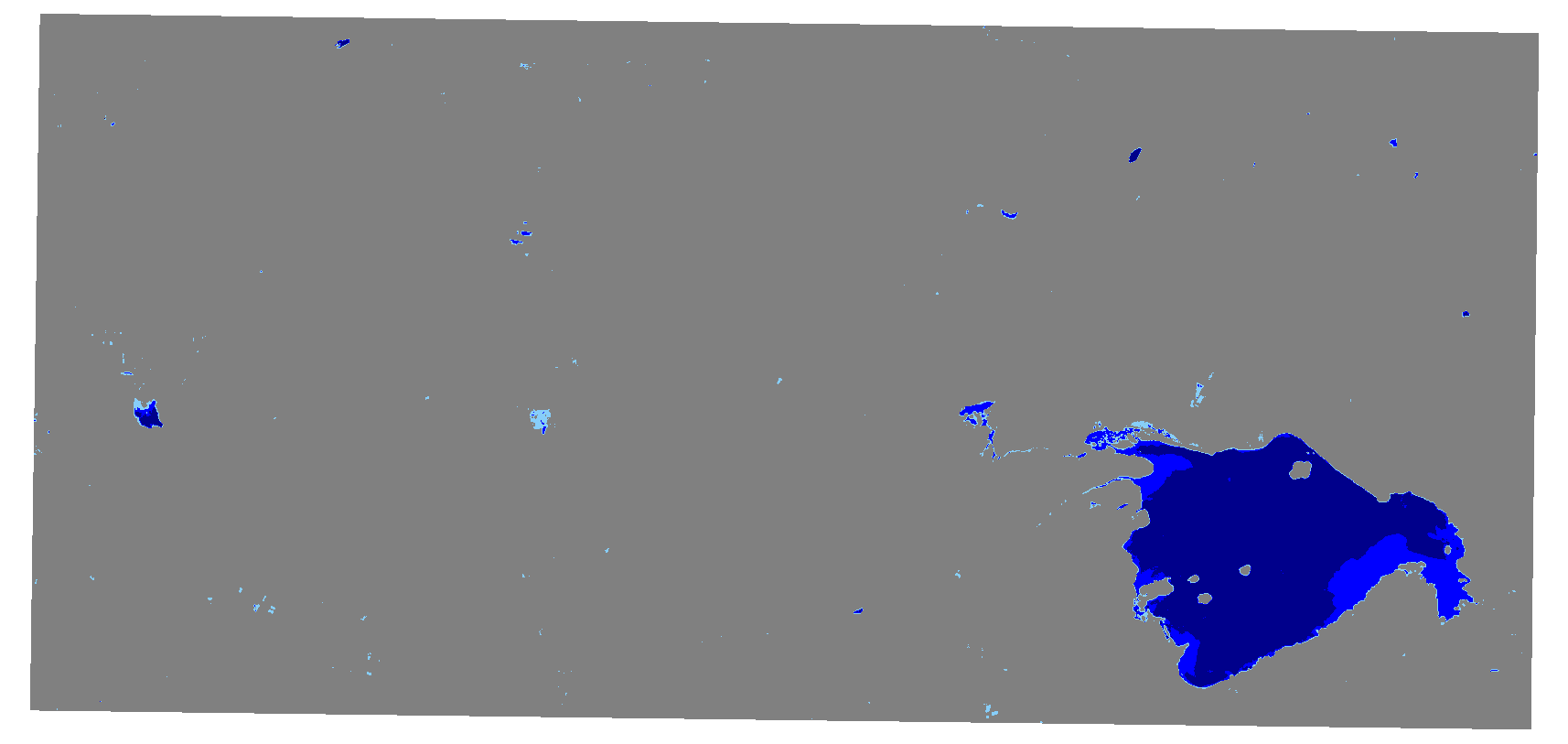}
         \caption{2024.}
         \label{fig:img5}
     \end{subfigure}
    \caption{Estimated percentage of water resources compared with the total surface of LandSat8 images for the Lake Cuitzeo (Michoacán, Mexico).}
    \label{fig:ndwi}
\end{figure}

\begin{figure}[t]
    \centering
    \begin{subfigure}[b]{0.49\textwidth}
         \centering
         \includegraphics[width=\textwidth]{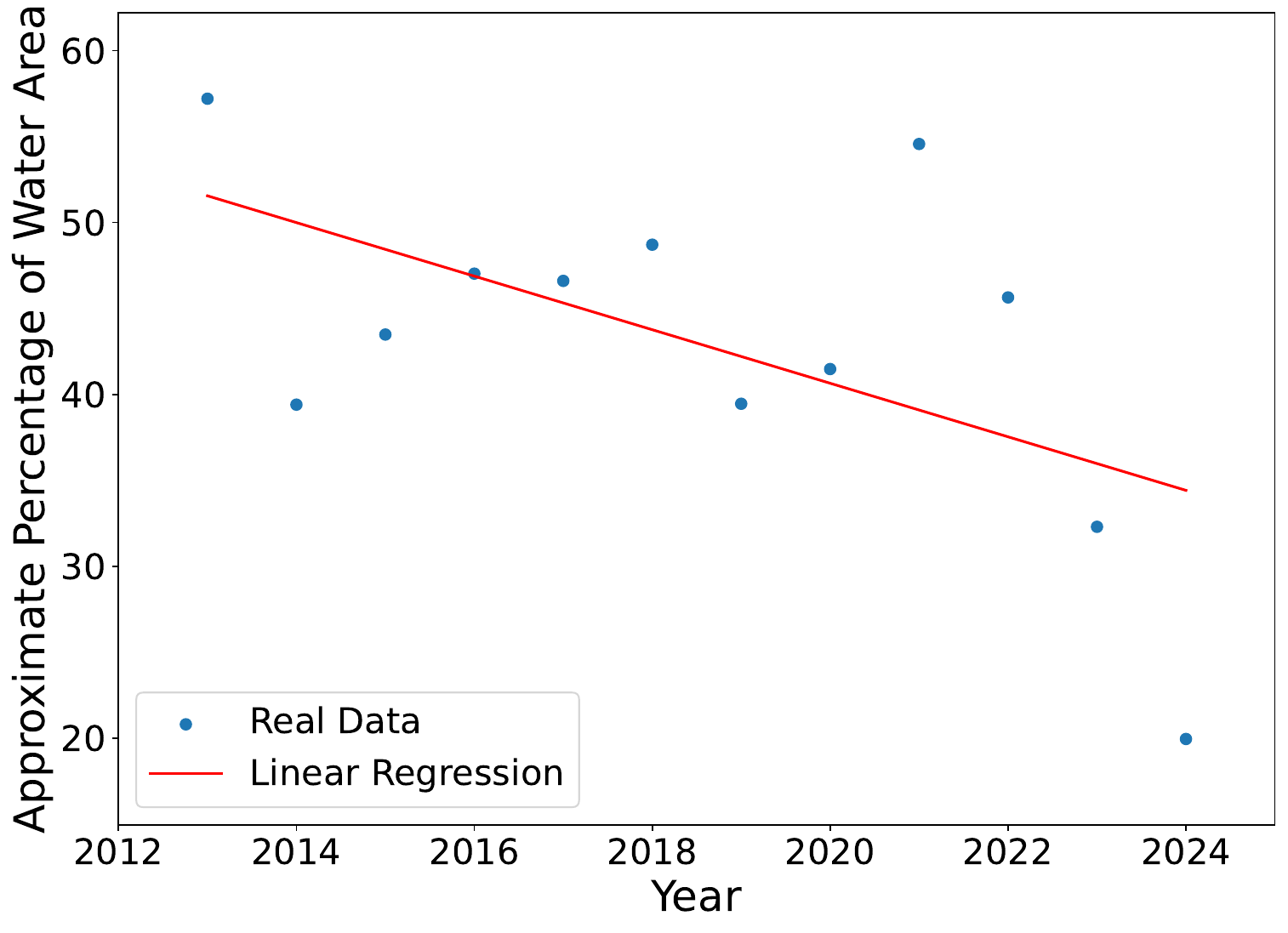}
         \caption{Path 27, Row 46.}
         \label{fig:path27row46}
     \end{subfigure}
     \hfill
     \begin{subfigure}[b]{0.49\textwidth}
         \centering
         \includegraphics[width=\textwidth]{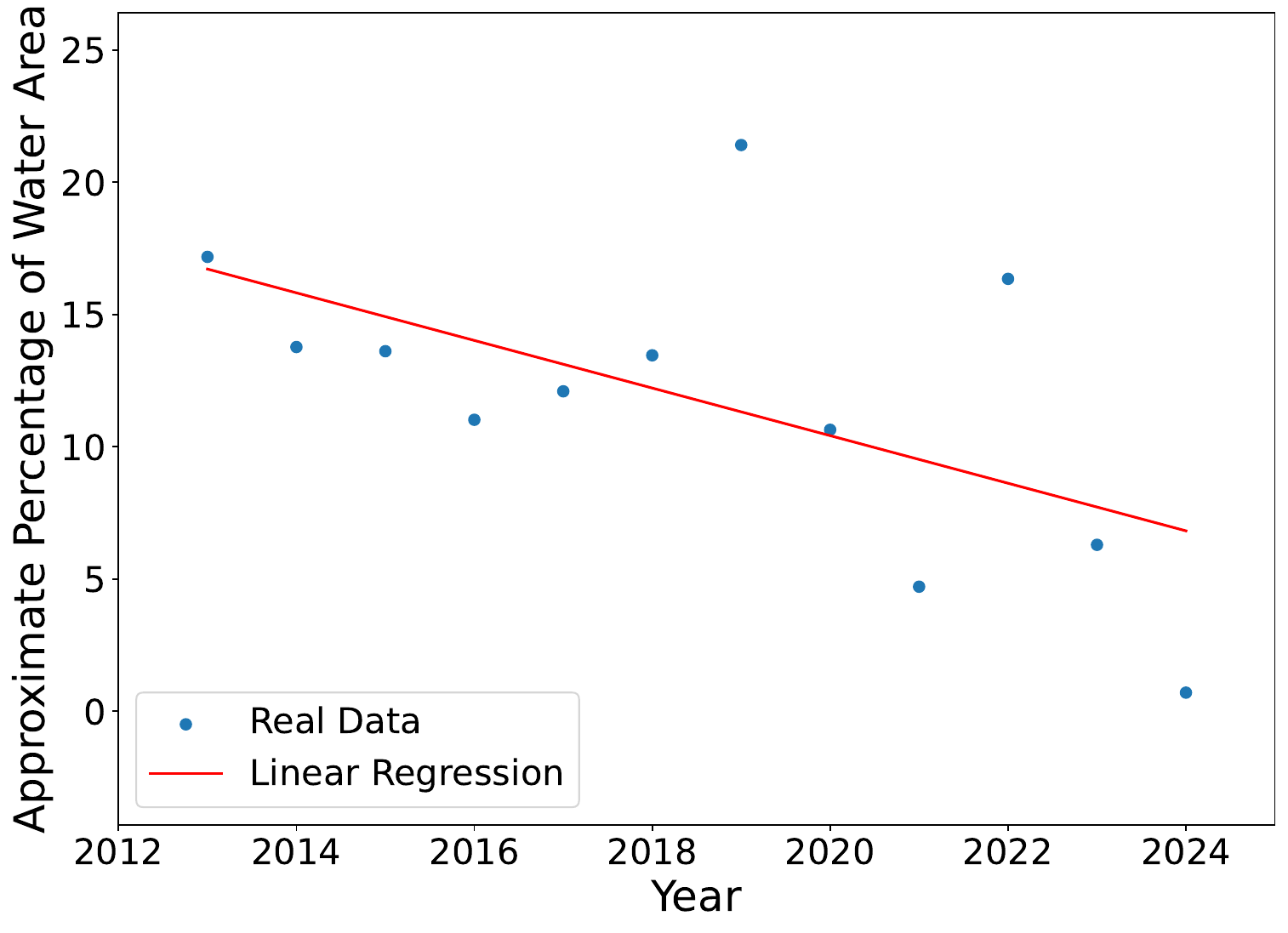}
         \caption{Path 28, Row 46.}
         \label{fig:path28row46}
     \end{subfigure}
    \caption{Estimated percentage of water resources compared with the total surface of LandSat8 images for the Lake Cuitzeo (Michoacán, Mexico).}
    \label{fig:summaries}
\end{figure}

Figure \ref{fig:casestudy} depicts the system design used for this case study. This system is composed of the following stages:
\begin{itemize}
    \item \textbf{Downloader}. This stage implements a function to download LandSat8 images from EarthExplorer \cite{earthexplorer}. It receives as input a spatio-temporal query to download the images. We performed a query with the coordinates (19.936739, -101.136399) for the range of data 2013-2024.
    \item \textbf{Decompressing}. It unpacks the downloaded images and makes the bands of the image available for the next stages.
    \item \textbf{Indexing}. This stage indexes the image metadata into a Geoportal.
    \item \textbf{Cropping}. It crops from each band a surface corresponding to a bounding box corresponding to the Lake Cuitzeo.
    \item \textbf{Derivates}. It generates NDWIs using different band combinations as described by Yan et al. \cite{yan2017analysis}. Figure \ref{fig:ndwi} shows some examples of the outputs produced by this stage.
    \item \textbf{Summary}. This stage summarizes the percentage of water identified by the indexes. Figure \ref{fig:summaries} shows examples of the output of this stage.
\end{itemize}

\begin{table}[t]
\caption{Characteristics of the infrastructure used for evaluation.} \label{tb:infraestructure}
\centering
\begin{tabular}{lllllll}
\toprule[0.3mm]
\textbf{Label} &   & \textbf{Storage Capacity (TB)} &  & \textbf{Number of cores} &  & \textbf{RAM (GB)} \\ \hline
gamma & & 1.8 & & 48 & & 126 \\
alpha & & 11 & & 48 & & 126 \\
disys18 & & 2.7 & & 24 & & 252 \\
\bottomrule[0.3mm]
\end{tabular}
\end{table}

We use a dataset of 213 LandSat8 images with a total size of 251 GB as input. We deployed the system depicted in Figure \ref{fig:casestudy} in a private cloud in Mexico. Table \ref{tb:infraestructure} shows the infrastructure details of the machines used. The idea was to replicate a scenario where different consumer and producer machines transfer and process data. Thus, the data downloader and decompressing stages were deployed on alpha, whereas we deployed the indexing, cropping, and derivates stages on gamma. Finally, we deployed the summary and Geoportal stages on disys18.

\begin{figure}[t]
    \centering
    \begin{subfigure}[b]{0.49\textwidth}
         \centering
         \includegraphics[width=\textwidth]{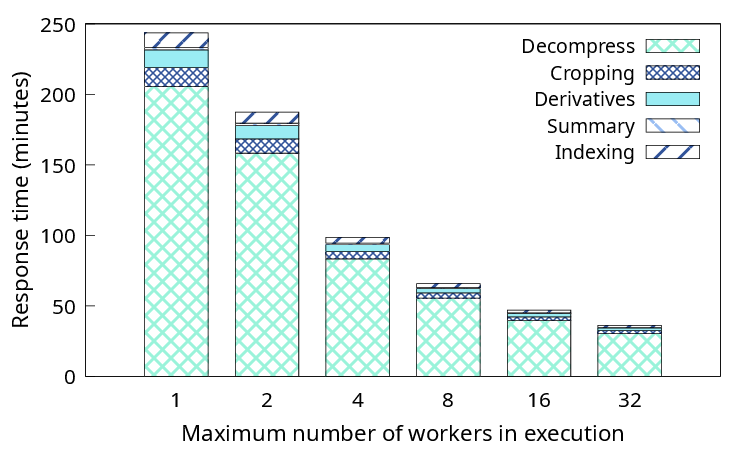}
         \caption{32 images.}
         \label{fig:times1}
     \end{subfigure}
     \hfill
     \begin{subfigure}[b]{0.49\textwidth}
         \centering
         \includegraphics[width=\textwidth]{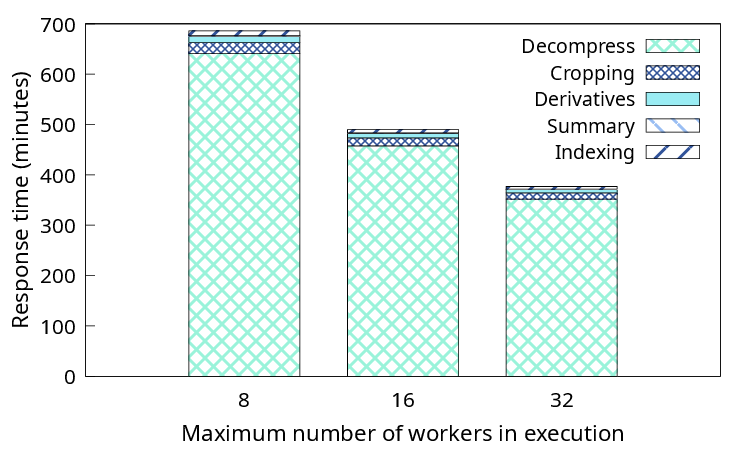}
         \caption{213 images.}
         \label{fig:times2}
     \end{subfigure}
    \caption{Response time observed when processing 32 and 213 LanSat8 images.}
    \label{fig:times}
\end{figure}

First, we evaluate the performance of GeoNimbus to manage data processing through the system depicted in Figure \ref{fig:casestudy}. Thus, we process 32 LandSat8 images (39 GB) using various maximum parallel workers. Figure \ref{fig:times1} shows on the vertical axis the response time observed for a varying number of workers (horizontal axis). As can be observed, the increment in the number of workers reduces the response time, which is an expected behavior in parallel systems. The 32 images were processed in 245.45 minutes using only one worker, whereas with 32 workers, the processing was completed in 36.10 minutes. This means an improvement in the response time of 85.10\%.

Figure \ref{fig:times2} shows, in the vertical axis, the response time observed to process the 213 images with 8, 16, and 32 workers. This experiment aims to show how GeoNimbus scales when managing large workloads. Again, we observe a reduction in the response times when more workers are added. With 32 workers, the complete dataset (251 GB) was processed in 6.28 hours, whereas with eight workers in 11.43 hours. This means an acceleration of 1.82x. 

Figure \ref{fig:ndwi} shows examples of NDWI$_{red}$ produced with a combination of the bands 7 (SWIR 2) and 4 (red). We present results from 2013  to 2024. These images correspond to a region with a path=27 and a row=46. Using the visual information generated by the NDWI, we can observe that the left side of the lake almost disappeared in 2024 (Figure \ref{fig:img5}). 

Figure \ref{fig:summaries} shows results obtained for images with paths 27 and 28 from the summary stage. Figures \ref{fig:path27row46} and \ref{fig:path28row46} show on the vertical axis the approximate percentage of water area in the index for different years from 2012 to 2024 (horizontal axis). In both Figures, the dots correspond to the real values obtained with the indexes, whereas the red line is a linear regression obtained to visualize trends in the real values. We calculated the approximate percentage of water area by counting the number of pixels in the image with a value higher than an umbral number. Those pixels higher than this umbral are tagged as water (blue zones in Figure \ref{fig:path28row46}), and the percentage is obtained by dividing the number of water pixels by the total number of pixels. For these experiments, we determine an umbral number of $0.65$. Both figures show a trend where water resources have decreased over the last four years (2021-2024). For example, in Figures \ref{fig:path27row46} (which correspond to the images shown in Figure \ref{fig:ndwi}), it can be observed that from 2021 to 2024, the percentage decreases from 57.20\% to 19.95\%. 

\section{Conclusions} \label{sec:conclusions}

In this paper, we presented GeoNimbus, a framework for designing EOS that follows the design principles of serverless computing. GeoNimbus automatically manages the deployment, scaling, monitoring, and execution of functions and applications through different infrastructures. One of GeoNimbus's goals is to reduce the complexity of creating serverless EOS and enable large-scale environmental studies.

To evaluate GeoNimbus, we conducted a case study based on the processing and analysis of LandSat8 images corresponding to Lake Cuitzeo. In terms of performance, the evaluation shows the scalability of GeoNimbus by parallelizing the stages in an EOS. GeoNimbus decreases by almost 85\% the time required to process the data in comparison with a non-parallel configuration. From the results of the EOS evaluated in this case study, we observe from the NDWI products generated that the water extension of Lake Cuitzeo has highly decreased in the past five years. In future work, we plan to perform other environmental studies to measure changes in vegetation and urban areas. We are also working on performing data fusion analysis using weather data acquired from ground stations with the results produced from the image indexes.

\section*{Acknowledgments}

The Spanish Research Agency has partially funded this work through the project ``New scalable I/O techniques for hybrid HPC and data-intensive workloads (SCIOT)'' with reference PID2022-138050NB-I00.

\bibliographystyle{splncs04}
\bibliography{references}

\begin{thebibliography}{10}
\providecommand{\url}[1]{\texttt{#1}}
\providecommand{\urlprefix}{URL }
\providecommand{\doi}[1]{https://doi.org/#1}

\bibitem{lambda}
AWS, A.: Aws lambda, \url{https://aws.amazon.com/lambda/}, [Accessed
  05-05-2024]

\bibitem{barron2024gis}
Barron-Lugo, J.A., Lopez-Arevalo, I., Gonzalez-Compean, J.,
  Alvarado-Barrientos, M.S., Carretero, J., Sosa-Sosa, V.J., Montella, R.: A
  gis-big data model for improving the coverage and analysis processes of
  territory observation, and integrating ground-based observations with
  retrospective meteorological data. International Journal of Applied Earth
  Observation and Geoinformation  \textbf{128},  103736 (2024)

\bibitem{CARRIZALESESPINOZA2024}
Carrizales-Espinoza, D., Sanchez-Gallegos, D.D., Gonzalez-Compean, J.,
  Carretero, J.: Structmesh: A storage framework for serverless computing
  continuum. Future Generation Computer Systems  (2024).
  \doi{https://doi.org/10.1016/j.future.2024.05.033},
  \url{https://www.sciencedirect.com/science/article/pii/S0167739X24002401}

\bibitem{chard2020funcx}
Chard, R., Babuji, Y., Li, Z., Skluzacek, T., Woodard, A., Blaiszik, B.,
  Foster, I., Chard, K.: Funcx: A federated function serving fabric for
  science. In: Proceedings of the 29th International symposium on
  high-performance parallel and distributed computing. pp. 65--76 (2020)

\bibitem{cloudfunctions}
Cloud, G.: Cloud functions, \url{https://cloud.google.com/functions}, [Accessed
  05-05-2024]

\bibitem{jonas2019cloud}
Jonas, E., Schleier-Smith, J., Sreekanti, V., Tsai, C.C., Khandelwal, A., Pu,
  Q., Shankar, V., Carreira, J., Krauth, K., Yadwadkar, N., et~al.: Cloud
  programming simplified: A berkeley view on serverless computing. arXiv
  preprint arXiv:1902.03383  (2019)

\bibitem{kaiser2021towards}
Kaiser, D., Dovhan, B., Bauer, A., Kounev, S.: Towards splitting monolithic
  workflows into serverless functions and estimating their run-time in the
  earth observation domain (poster). In: SSP (2021)

\bibitem{kurniawan2019introduction}
Kurniawan, A., Lau, W., Kurniawan, A., Lau, W.: Introduction to azure
  functions. Practical Azure Functions: A Guide to Web, Mobile, and IoT
  Applications pp. 1--21 (2019)

\bibitem{morales2018data}
Morales-Ferreira, P., Santiago-Duran, M., Gaytan-Diaz, C., Gonzalez-Compean,
  J., Sosa-Sosa, V.J., Lopez-Arevalo, I.: A data distribution service for cloud
  and containerized storage based on information dispersal. In: 2018 IEEE
  Symposium on Service-Oriented System Engineering (SOSE). pp. 86--95. IEEE
  (2018)

\bibitem{nativi2015big}
Nativi, S., Mazzetti, P., Santoro, M., Papeschi, F., Craglia, M., Ochiai, O.:
  Big data challenges in building the global earth observation system of
  systems. Environmental Modelling \& Software  \textbf{68},  1--26 (2015)

\bibitem{opara2014critical}
Opara-Martins, J., Sahandi, R., Tian, F.: Critical review of vendor lock-in and
  its impact on adoption of cloud computing. In: International conference on
  information society (i-Society 2014). pp. 92--97. IEEE (2014)

\bibitem{sanchez2022puzzlemesh}
Sanchez-Gallegos, D.D., Gonzalez-Compean, J., Carretero, J., Marin-Castro,
  H.M., Tchernykh, A., Montella, R.: Puzzlemesh: A puzzle model to build mesh
  of agnostic services for edge-fog-cloud. IEEE Transactions on Services
  Computing  \textbf{16}(2),  1334--1345 (2022)

\bibitem{ueckermann2020podpac}
Ueckermann, M.P., Bieszczad, J., Entekhabi, D., Shapiro, M.L., Callendar, D.R.,
  Sullivan, D., Milloy, J.: Podpac: open-source python software for enabling
  harmonized, plug-and-play processing of disparate earth observation data sets
  and seamless transition onto the serverless cloud by earth scientists. Earth
  Science Informatics  \textbf{13},  1507--1521 (2020)

\bibitem{ullah2023orchestration}
Ullah, A., Kiss, T., Kov{\'a}cs, J., Tusa, F., Deslauriers, J., Dagdeviren, H.,
  Arjun, R., Hamzeh, H.: Orchestration in the cloud-to-things compute
  continuum: taxonomy, survey and future directions. Journal of Cloud Computing
   \textbf{12}(1), ~135 (2023)

\bibitem{earthexplorer}
USGS: Earthexplorer - usgs (July 31, 2020),
  \url{https://earthexplorer.usgs.gov/s}

\bibitem{yan2017analysis}
Yan, D., Wang, X., Zhu, X., Huang, C., Li, W.: Analysis of the use of ndwigreen
  and ndwired for inland water mapping in the yellow river basin using
  landsat-8 oli imagery. Remote Sensing Letters  \textbf{8}(10),  996--1005
  (2017)

\end{thebibliography}

\end{document}